\documentclass[envcountsame]{llncs}

\usepackage{latexsym}
\newtheorem{ourclaim}[theorem]{Claim}
\newtheorem{ourconjecture}[theorem]{Conjecture}

\def\qed{{\ \nolinebreak\hfill\mbox{$\Box$}}\smallskip}	
\renewenvironment{proof}{\noindent {\bf Proof.}\quad}{\qed}

\newcommand{\ISO}{\mbox{\rm ISO}}

\newcommand{\GI}{\mbox{\rm GI}}

\newcommand{\cset}{\mbox{$\cal C$}}

\newcommand{\np}{\mbox{\rm NP}} %
\newcommand{\p}{\mbox{\rm P}}  %
\newcommand{\conp}{\mbox{\rm coNP}} %

\newcommand{\csp}{\mbox{\rm CSP}}

\newcommand{\iso}{\mbox{\rm ISO}}
\newcommand{\isoimp}{\mbox{\rm ISO-IMP}}
\newcommand{\threepart}{\mbox{\rm 3-Partition}}
\newcommand{\unarythreepart}{\mbox{\rm Unary-3-Partition}}

\newcommand{\sat}{\mbox{\rm SAT}}

\DeclareSymbolFont{AMSb}{U}{msb}{m}{n}
\DeclareMathSymbol{\N}{\mathbin}{AMSb}{"4E}
\DeclareMathSymbol{\Z}{\mathbin}{AMSb}{"5A}
\DeclareMathSymbol{\R}{\mathbin}{AMSb}{"52}
\DeclareMathSymbol{\Q}{\mathbin}{AMSb}{"51}

\newcommand{\isoimpl}{\widetilde{\Rightarrow}}
\newcommand{\parallelnp}{\mbox{$\p_{||}^{\mbox{\rm \scriptsize NP}}$}}

\title{Isomorphic Implication\thanks{%
Supported in part by grants NSF-CCR-0311021 and DFG VO 630/5-1.}}
\author{Michael Bauland\inst{1}
\and
Edith Hemaspaandra\inst{2}
}

\institute{Theoretische Informatik, Universit\"at Hannover, Appelstr.\ 4, 
D-30167 Hannover, Germany,
email: bauland@thi.uni-hannover.de. \\
Work done in part while visiting the
Laboratory for Applied Computing at
Rochester Institute of Technology.
\and
Department of Computer Science,
Rochester Institute of Technology,
Rochester, NY 14623, U.S.A.,
e-mail: eh@cs.rit.edu. \\
Work done in part while on sabbatical at the
University of Rochester.
}

\begin{document}

\pagestyle{plain}

\maketitle

\begin{abstract}
We study the isomorphic implication problem for Boolean constraints.
We show that this is a natural analog of the subgraph isomorphism
problem.  We prove that, depending on the set of constraints,
this problem is in P, NP-complete, or NP-hard, coNP-hard,
and in $\parallelnp$.  We show how to extend the NP-hardness and
coNP-hardness to $\parallelnp$-hardness for some cases,
and conjecture that this can be done in all cases.
\end{abstract}

\section{Introduction}
\label{s:introduction}
One of the most interesting and well-studied problems in complexity
theory is the graph isomorphism problem (GI). This is the problem
of determining whether two graphs are isomorphic, i.e., whether
there exists a renaming of vertices such that the graphs become 
equal.  This is a fascinating
problem, since it is the most natural example of a problem that is
in NP, not known to be in P, and unlikely to be
NP-complete (see~\cite{koe-sch-tor:b:graph-iso}).

The obvious analog of graph isomorphism for Boolean formulas
is the formula isomorphism problem.  This is the problem of
determining whether two formulas are isomorphic, i.e., whether
we can rename the variables such that the formulas become
equivalent.  This problem has the same behavior as the graph
isomorphism problem one level higher in the polynomial hierarchy:
The formula isomorphism problem is in $\Sigma_2^p$, NP-hard, and unlikely to be
$\Sigma_2^p$-complete~\cite{agr-thi:j:boolean-isomorphism}.

Note that graph isomorphism can be viewed as a special case of Boolean
isomorphism, since graph isomorphism corresponds to Boolean
isomorphism for 2-positive-CNF formulas, in the following way:
Every graph $G$ (without isolated vertices) corresponds to the (unique)
formula $\bigwedge_{\{i,j\} \in E(G)} x_i \vee x_j$.  Then two graphs
without isolated vertices are isomorphic if and only if their
corresponding formulas are isomorphic.

One might wonder what happens when we look at other restrictions
on the set of formulas.
There are general frameworks for looking at all restrictions on Boolean
formulas: The most often used is the Boolean constraint framework
introduced by Schaefer~\cite{sch:c:sat}.
Basically (formal definitions can be found
in the next section) we look at formulas as CNF formulas
(or sets of clauses) where each clause is an application of a 
constraint (a $k$-ary Boolean function) to a list of variables. 
Each finite set of constraints gives rise to a new language, and
so there are an infinite number of languages to consider.  Schaefer
studied the satisfiability problem for all finite sets of constraints.
He showed that all of these satisfiability problems are either
in P or NP-complete, and he gave a simple criterion to determine
which of the cases holds.

The last decade has seen renewed interest in Schaefer's result,
and has seen many dichotomy (and dichotomy-like) theorems for
problems related to the satisfiability of Boolean constraints.
For example, such results were obtained for
the maximum satisfiability problem~\cite{cre:j:max-sat},
counting satisfying assignments~\cite{cre-heb:j:counting},
the inverse satisfiability problem~\cite{kav-sid:j:inv-sat},
the unique satisfiability problem~\cite{jub:c:usat},
the minimal satisfying assignment problem~\cite{kir-kol:c:minimal-sat},
approximability problems~\cite{kha-sud-tre-wil:j:approx-constraints}, 
and the equivalence problem~\cite{boe-hem-rei-vol:c:constraints}.
For an excellent survey of dichotomy theorems for Boolean
constraint satisfaction problems, see~\cite{cre-kha-sud:b:con}.

Most of the results listed above were proved using methods similar
to the one used by Schaefer~\cite{sch:c:sat}.
A more recent approach to proving results of this form is
with the help of the so-called algebraic
approach~\cite{jea:j:algebraic-structure,jea-coh-gys:j:closure-properties-constraints,bul-jea-kro:c:finite-algebras}.
This approach uses the 
clone (closed classes) structure of Boolean functions called Post's lattice,
after Emil Post, who first identified these classes \cite{pos:j:re}. A good 
introduction of how this can be used to obtain short proofs
can be found in \cite{boe-cre-rei-vol:j:blocks-two}.
However, this approach does not work for isomorphism problems,
because it uses existential quantification.

For the case of most interest for this paper, the Boolean isomorphism 
problem for constraints,
B\"ohler et al.~\cite{boe-hem-rei-vol:c:constraints,boe-hem-rei-vol:c:iso,boe-hem-rei-vol:t:con-iso-revised}
have shown that this
problem is in P, GI-complete, or GI-hard, coNP-hard, and in
$\parallelnp$ (the class of problems solvable in polynomial
time with one round of parallel queries to NP).  As in Schaefer's
theorem, simple properties of the set of constraints determine
the complexity.

A problem closely related to the graph isomorphism problem is the
subgraph isomorphism problem.  This is the problem, given
two graphs $G$ and $H$, to determine whether $G$ contains a
subgraph isomorphic to $H$.  In contrast to the graph isomorphism
problem, the subgraph isomorphism problem can easily be seen
to be NP-complete (it contains, for example, CLIQUE, HAMILTONIAN CYCLE,
and HAMILTONIAN PATH). 

To further study the relationship between the isomorphism problems
for graphs and constraints, we would like to find a relation ${\cal R}$
on constraints that is to isomorphism for constraints as the
subgraph isomorphism problem is to graph isomorphism.

Such a relation ${\cal R}$ should at least have the following properties:
\begin{enumerate}
\item
A graph $G$ is isomorphic to a graph $H$ if and 
only if $G$ contains a subgraph isomorphic to $H$ and $H$ contains
a subgraph isomorphic to $G$.  We want the same property in the constraint
case, i.e.,
for $S$ and $U$ sets of constraint applications,
$S$ is isomorphic to $U$ if and only if
$S {\cal R} U$ and $U {\cal R} S$.
\item
The subgraph isomorphism problem should be a special case
of the decision problem induced by ${\cal R}$, in the same way as the graph isomorphism
problem is a special case of the constraint isomorphism 
problem.  In particular, for $G$ and $H$ graphs, 
let $S(G)$ and $S(H)$ be
their (standard) translations into sets of constraint applications
of $\lambda xy.(x \vee y)$, i.e., $S(G) =
\{x_i \vee x_j \ | \ \{x_i,x_j\} \in E(G)\}$ and
$S(H) =
\{x_i \vee x_j \ | \ \{x_i,x_j\} \in E(H)\}$.
For $G$ and $H$ graphs without isolated vertices,
$G$ is isomorphic to $H$ if and only if $S(G)$ is
isomorphic to $S(H)$. 
We want $G$ to have a subgraph isomorphic to $H$ if and only
if $S(G) {\cal R} S(H)$.
\end{enumerate}

Borchert et al.~\cite[p.~692]{bor-ran-ste:j:boolean-equivalence}
suggest using the subfunction relations $\gg_v$ and
$\gg_{cv}$ as analogs of subgraph isomorphism.  These
relations are defined as follows.
For two formulas $\phi$ and $\psi$, $\phi \gg_v \psi$ if and only
if there exists a function $\pi$ from
variables to variables such that $\pi(\phi)$ is equivalent to
$\psi$.
$\phi \gg_{cv} \psi$ if and only if there exists a function $\pi$ from
variables to variables and constants such that $\pi(\phi)$ is equivalent to
$\psi$~\cite{bor-ran:t:subfunction}.
Borchert and Ranjan~\cite{bor-ran:t:subfunction}
show that these relations satisfy our first desirable property,
i.e.,  $S$ is isomorphic to $U$ if and
only if $S \gg_v U$ and $U \gg_v S$, and
that $S$ is isomorphic to $U$ if and
only if $S \gg_{cv} U$ and $U \gg_{cv} S$.
They also show that the problem of determining whether
$\phi \gg_v \psi$ and the problem of
determining whether $\phi \gg_{cv} \psi$,
for unrestricted Boolean formulas, are $\Sigma_2^p$-complete.

But  Borchert et al.'s subfunction relations will not give the
second desirable property.  Consider, for example, the
graphs $G$ and $H$ such that $V(G) = V(H) = \{1,2,3\}$,
$E(G) = \{\{1,2\}, \{1, 3\}, \{2, 3\}\}$, and 
$E(H) = \{\{1,2\}, \{1, 3\}\}$.  Clearly, $G$ contains a subgraph isomorphic
to $H$, but $(x_1 \vee x_2) \wedge (x_1 \vee x_3) \wedge (x_2 \vee x_3)
\not \gg_{cv} (x_1 \vee x_2) \wedge (x_1 \vee x_3)$.

How could the concept of a subgraph be translated to sets of
constraint applications?
As a first attempt at translating subgraph isomorphism to
constraint isomorphism one might try the following: For sets of
constraint applications $S$ and $U$, does there exist a subset
$\widehat{S}$ of $S$ that is isomorphic to $U$. Certainly, such a definition
satisfies the second desired property.  But this definition
does not satisfy the first desired property, since
it is quite possible for sets of constraint applications
to be equivalent without being equal.

We claim that isomorphic implication satisfies both
desired properties, and is a natural analog of the subgraph isomorphism
problem for Boolean constraints.

For $S$ and $U$ sets of constraint applications over variables
$X$, we say that $S$ isomorphically implies $U$
(notation: $S \isoimpl U$)
if and only if there exists
a permutation $\pi$ on $X$ such that $\pi(S) \Rightarrow U$.
In Section~\ref{s:complexity},
we show that, depending on the set of constraints,
the isomorphic implication problem 
is in P, NP-complete, or NP-hard, coNP-hard, and in $\parallelnp$.
Our belief is that the isomorphic implication problem is
$\parallelnp$-complete for all the cases where it is both NP-hard and
coNP-hard. In Section~\ref{s:conjecture}, we prove this conjecture for
some cases.  

\section{Preliminaries}
\label{s:preliminaries}

We will mostly use the constraint terminology from~\cite{cre-kha-sud:b:con}.

\begin{definition}
\begin{enumerate}
\item A \emph{constraint} $C$ (of arity $k$) is a Boolean function from 
$\{0,1\}^k$ to $\{0,1\}$.
\item If $C$ is a constraint of arity $k$, and $z_1,z_2,\ldots,z_k$ are (not
necessarily distinct) variables, then $C(z_1,z_2,\ldots,z_k)$ is a 
\emph{constraint application of} $C$.
\item If $C$ is a constraint of arity $k$, and for $1 \leq i \leq k$, $z_i$ is
a variable or a constant (0 or 1), then $C(z_1,z_2,\ldots,z_k)$ is a constraint
application of $C$ \emph{with constants}.
\item If $S$ is a set of constraint applications [with constants]
and $X$ is a set of variables that includes all variables that occur
in $S$, we  say that $S$ is a set of constraint applications 
[with constants] \emph{over variables $X$}.
\end{enumerate}
\end{definition}

\begin{definition}
Let $C$ be a $k$-ary constraint.
\begin{itemize}
\item $C$ is \emph{0-valid} if $C(0,\ldots,0)=1$.
\item $C$ is \emph{1-valid} if $C(1,\ldots,1)=1$.
\item $C$ is \emph{Horn} (or weakly negative) if $C(x_1, \ldots, x_k)$
is equivalent to a CNF
formula where each clause has at most one positive literal.
\item $C$ is \emph{anti-Horn} (or weakly positive) if $C(x_1, \ldots, x_k)$
is equivalent to a
CNF formula where each clause has at most one negative literal.
\item $C$ is \emph{bijunctive} if $C(x_1, \ldots, x_k)$
is equivalent to a 2CNF formula.
\item $C$ is \emph{affine} if $C(x_1, \ldots, x_k)$
is equivalent to an XOR-CNF formula.
\item $C$ is \emph{2-affine} (or affine of width 2) if $C(x_1, \ldots, x_k)$
is equivalent to an
XOR-CNF formula, such that every clause contains at most two literals.
\item $C$ is \emph{complementive} (or C-closed) if for every $s\in \{0,1\}^k$,
 $C(s)=C(\overline{s})$, where $\overline{s}\in 
 \{0,1\}^k =_{def} (1,\ldots,1)-s$, i.e., $\overline{s}$ is obtained by
 flipping every bit of $s$.
\end{itemize}
\end{definition}

Let \cset\ be a finite set of constraints.
We say \cset\ is 
0-valid, 1-valid, Horn, anti-Horn, bijunctive, affine, 2-affine, or 
complementive, if \emph{every}
constraint $C\in\cset$ has this respective property.
We say that \cset\ is
\emph{Schaefer} if \cset\ is Horn, anti-Horn, affine, or bijunctive.

\begin{definition}[\cite{boe-hem-rei-vol:c:constraints}]
Let \cset\ be a finite set of constraints.
\begin{enumerate}
\item $\iso(\cset)$ is the problem, given two sets $S$ and $U$ of 
constraint applications of \cset\ over variables $X$,
to decide whether
$S$ is isomorphic to $U$ (denoted by $S \cong U$), 
i.e., whether there exists a permutation $\pi$ of $X$ such that
$\pi(S) \equiv U$;
Here $\pi(S)$ is the set of constraint applications that results
when we simultaneously replace every variable $x$ in $S$ by
$\pi(x)$.
\item $\iso_c(\cset)$ is the problem, given two sets $S$ and $U$ of 
constraint applications of \cset\ with constants, to decide whether $S$ 
is isomorphic to $U$.
\end{enumerate}
\end{definition}

\begin{theorem}[\cite{boe-hem-rei-vol:c:constraints,boe-hem-rei-vol:c:iso,boe-hem-rei-vol:t:con-iso-revised}]
\label{t:isotrich}
Let ${\cal C}$ be a finite set of constraints.
\begin{enumerate}
\item If ${\cal C}$ is not Schaefer, then
$\ISO({\cal C})$ and $\ISO_c({\cal C})$ are {\rm coNP}-hard,
\GI-hard, and in $\parallelnp$.
\item If ${\cal C}$ is Schaefer and not 2-affine,
then $\ISO({\cal C})$ and $\ISO_c({\cal C})$ are polynomial-time
many-one equivalent to \GI.
\item Otherwise, ${\cal C}$ is 2-affine and
$\ISO({\cal C})$ and $\ISO_c({\cal C})$ are in {\rm P}.
\end{enumerate}
\end{theorem}

The isomorphic implication problem combines isomorphism with
implication in the following way.

\begin{definition}
\label{d:isoimp}
Let \cset\ be a finite set of constraints.
\begin{enumerate}
\item $\isoimp(\cset)$ is the problem, given two sets $S$ and $U$ of 
constraint applications of \cset\ over variables $X$, to decide whether
$S$ isomorphically implies $U$ (denoted by $S \isoimpl U$),
i.e., whether there exists a permutation $\pi$ of $X$ such that
$\pi(S) \Rightarrow U$;
Here $\pi(S)$ is the set of constraint applications that results
when we simultaneously replace every variable $x$ in $S$ by
$\pi(x)$.

\item $\isoimp_c(\cset)$ is the problem, given two sets $S$ and $U$ of
constraint applications of \cset\ with constants, deciding whether $S$
isomorphically implies $U$.
\end{enumerate}
\end{definition}

To show that this definition is well defined
we need to show that if $S$ and $U$ are sets of constraint applications over
variables $X$, $Y$ is a set of variables disjoint from
$X$, and there exists a permutation $\pi$ of $X \cup Y$ such that
$\pi(S) \Rightarrow U$, then there exists a permutation
$\rho$ of $X$ such that $\rho(S) \Rightarrow U$.

Suppose that $\pi$ is a permutation of $X \cup Y$
such that $\pi(S) \Rightarrow U$ and $||\{y \in Y \  | \ \pi(y) \in X\}||$
is minimal and at least one.  Let $x,x' \in X$ and $y,y' \in Y$ be such
that $\pi(x') = y$ and $\pi(y') = x$.  Define a new permutation
$\rho$ as follows: $\rho(x') = x$,
$\rho(y') = y$, and $\rho(z) = \pi(z)$ for all $z \in (X \cup Y) - \{x',y'\}$. 
We will show that $\rho(S) \Rightarrow U$.  This is a contradiction,
since $||\{y \in Y \  | \ \rho(y) \in X\}|| = 
||\{y \in Y \  | \ \pi(y) \in X\}|| - 1$.

Let $Z$ be a list of the variables in $(X \cup Y) - \{x,y\}$.
Suppose that $\rho(S)(Z,x,y) \not \Rightarrow U(Z,x,y)$.  Then there exists 
a string $s \in \{0,1\}^{||Z||}$ and $a, b \in \{0,1\}$ such that
$\rho(S)(s,a,b) = 1$ and $U(s,a,b) = 0$.
Since $y$ does not occur in $U$, $U(s,a,\overline{b}) = 0$.
Since $\pi(S)(Z,x,y) \Rightarrow U(Z,x,y)$, it follows
that $\pi(S)(s,a,b) = 0$ and $\pi(S)(s,a,\overline{b}) = 0$.
Since $y'$ does not occur in $S$, $x$ does not occur in $\pi(S)$,
and so $\pi(S)(s,\overline{a},b) = 0$ and
$\pi(S)(s,\overline{a},\overline{b}) = 0$.  It follows that
$\pi(S)(s,x,y) \equiv 0$.  But note that
$\pi(S)(s,b,a) = \rho(S)(s,a,b) = 1$.  This is
a contradiction.

\begin{definition}
\begin{enumerate}
\item
The {\em graph isomorphism problem} is the problem, given
two graphs $G$ and $H$, to decide whether $G$ and $H$ are isomorphic,
i.e., whether there exists
a bijection $\pi$ from $V(G)$ to $V(H)$ such that
$\pi(G) = H$.  $\pi(G)$ is the graph such that $V(\pi(G)) = 
\{\pi(v) \ | \ v \in V(G) \}$ and 
$E(\pi(G)) = \{\{\pi(v), \pi(w)\} \ | \ \{v,w\} \in E(G) \}$.
\item 
The subgraph isomorphism problem is the problem, given two graphs $G$
and $H$, to decide whether $G$ contains a subgraph isomorphic to $H$, i.e.,
whether there exists a graph $G'$ such that $V(G') \subseteq V(G)$ and
$E(G') \subseteq E(G)$ and $G'$ is isomorphic to $H$.
\end{enumerate}
\end{definition}

\begin{theorem}[\cite{gar-joh:b:int,coo:c:theorem-proving}]
The subgraph isomorphism problem is \np-complete.
\end{theorem}

\begin{corollary}
The subgraph isomorphism problem for graphs without isolated
vertices is \np-complete.
\end{corollary}

\section{Subgraph Isomorphism and Isomorphic Implication}
\label{s:analog}

We will now show that the isomorphic implication problem
is a natural analog of the subgraph isomorphism problem,
in the sense explained in the introduction.

\begin{lemma}
\label{l:analog}
\begin{enumerate}
\item Let $S$ and $U$ be sets of constraint applications of
${\cal C}$ with constants.  Then $S \cong U$ if and
only if $S \isoimpl U$ and $U \isoimpl S$.
\item
For graphs $G$ and $H$ without isolated vertices, $G$ contains
a subgraph isomorphic to $H$ if and only if $S(G) \isoimpl S(H)$,
where $S$ is the ``standard'' translation from graphs to sets of
constraint applications of $\lambda xy. x \vee y$, i.e.,
for $\widehat{G}$ a graph,
$S(\widehat{G}) = \{ x_i \vee x_j \ | \ \{i,j\} \in E(\widehat{G})\}$.

\end{enumerate}
\end{lemma}

\begin{proof}
\begin{enumerate}
\item We claim that $S \cong U$ if and only if $S \isoimpl U$ and $U
\isoimpl S$.  The left-to-right direction is immediate.  For the converse,
let $X$ be the set of variables that occur in $S \cup U$.
Suppose that $\pi$ is a permutation of the variables occurring
in $S \cup U$ such that $\pi(S) \Rightarrow
U$ and that $\rho$ is a permutation of $X$ such that
$\rho(U) \Rightarrow S$.  Suppose for a contradiction that
$\pi(S) \not \equiv U$.  Then there exists an assignment
that satisfies $U$, and that does not satisfy $\pi(S)$.
Since $\rho(U) \Rightarrow S$, there are at least as many
satisfying assignments for $\pi(S)$ as for $U$.  It follows
that there exists an assignment that satisfies $\pi(S)$ and not $U$.
But that contradicts the assumption that $\pi(S) \Rightarrow U$.

\item 
Let $G$ and $H$ be graphs without isolated vertices. 
We will show that $G$ contains
a subgraph isomorphic to $H$ if and only if $S(G) \isoimpl S(H)$.

For the left-to-right direction,
let $G'$ be a subgraph of $G$ such that $G' \cong H$.
Let $\pi$ be a bijection from the vertices
of $G'$ to the vertices of $H$ such that $\pi(G') = H$.
Let $\rho$ be a permutation of the variables occurring
in $S(G) \cup S(H)$ such that $\rho(x_i) = x_{\pi(i)}$ for
all $i \in V(G')$.  It is easy to see that
$\rho(S(G')) = S(\pi(G')) = S(H)$.  Since $G'$ is 
a subgraph of $G$, $S(G') \subseteq S(G)$.
It follows that $S(H) \subseteq \rho(S(G))$, and thus,
$\rho(S(G)) \Rightarrow S(H)$.

For the converse, suppose that $S(G) \isoimpl S(H)$.  Let $\rho$ be
a permutation on the variables occurring in $S(G) \cup S(H)$ such that
$\rho(S(G)) \Rightarrow S(H)$.  It is easy to see that if
$\rho(S(G)) \Rightarrow x_i \vee x_j$, then
$x_i \vee x_j \in \rho(S(G))$. 
It follows that $S(H) \subseteq \rho(S(G))$.
Let $G'$ be such that $S(H) = \rho(S(G'))$
and $G'$ does not have isolated vertices (take $G' =
\pi^{-1}(H)$).
Since $\rho(S(G')) \subseteq \rho(S(G))$, it follows that
$S(G') \subseteq S(G)$, and thus, $G'$ is a subgraph of $G$.
Since $S(H) = \rho(S(G'))$ and $H$ and $G'$ do not
contain isolated vertices, it follows that $G'$ is isomorphic to $H$.
\end{enumerate}
\end{proof}

\section{Complexity of the Isomorphic Implication Problem}
\label{s:complexity}

The following theorem gives a trichotomy-like theorem for
the isomorphic implication problem.

\begin{theorem}\label{t:isoimp}
Let \cset\ be a finite set of constraints.
\begin{enumerate}
\item If every constraint in \cset\ is equivalent to a constant or
a conjunction of literals,
then $\isoimp(\cset)$ and $\isoimp_c(\cset)$ are in \p.
\item Otherwise, if \cset\ is Schaefer, then $\isoimp(\cset)$ and
$\isoimp_c(\cset)$ are \np-complete.
\item If \cset\ is not Schaefer, then $\isoimp(\cset)$ and
$\isoimp_c(\cset)$ are \np-hard,  \conp-hard, and in $\parallelnp$.
\end{enumerate}
\end{theorem}

\subsection{Upper bounds}
\label{s:upperbounds}

The NP upper bound for sets of constraints that are Schaefer is easy to see.
\begin{ourclaim}\label{clNP}
If \cset\ is Schaefer, then $\isoimp_c(\cset)$ is in \np.
\end{ourclaim}

\begin{proof}
Let $S$ and $U$ be sets of constraint applications of \cset\
with constants over variables $X$.
Then $S \isoimpl U$ if and only if there exists a permutation
$\pi$ of $X$ such that $\pi(S) \Rightarrow U$. Clearly, $\pi(S) \Rightarrow U$
if and only if $\pi(S) \cup U \equiv \pi(S)$.  Since \cset\ is Schaefer,
it can be determined in polynomial time whether two sets of
constraint applications of \cset\ with constants
are equivalent~\cite[Theorem 6]{boe-hem-rei-vol:c:constraints}.
\end{proof}

\begin{ourclaim}\label{clparallelNP}
For any finite set \cset\  of constraints,
$\isoimp_c(\cset)$ is in $\parallelnp$.
\end{ourclaim}

\begin{proof}(Similar to the argument before
\cite[Corollary 23]{boe-hem-rei-vol:c:constraints}.)
Let $S$ and $U$ be sets of constraint applications of ${\cal C}$ with 
constants.
Let $X$ be the set of all variables that occur in $S \cup U$.
{From} \cite[proof of Claim 22]{boe-hem-rei-vol:c:constraints},
we know that we can in polynomial
time with parallel access to NP compute
the set of all constraint applications of ${\cal C}$ with
constants over $X$ that are implied by $S$.  Call this
set $\widehat{S}$.
It is easy to see that $\pi(S) \Rightarrow U$ if and only if
$U \subseteq \widehat{S}$. It takes one query to
NP to find out whether there exists such a permutation. Since
two rounds of queries to NP are the same as one round of
queries to NP~\cite{bus-hay:j:tt}
it follows that we can determine whether $S \isoimpl U$ in $\parallelnp$.
\end{proof}

\begin{ourclaim}\label{clP}
Let \cset\ be a finite set of constraints such that every
constraint is equivalent to a constant or to a conjunction
of literals.  Then  $\isoimp_c(\cset)$ is in \p.
\end{ourclaim}

\begin{proof}
Let $S$ and $U$ be sets of constraint applications of ${\cal C}$
with constants.  We will view $S$ and $U$ as sets of literals and
constants.  
We first consider the case where $S$ or $U$ is equivalent
to a constant.  Note that it is easy to check if a set $X$ of literals
and constants is equivalent to $0$ or $1$, since $X$ is equivalent
to 1 if and only if $X = \{1\}$, and
$X$ is equivalent to 0 if and only if $0 \in X$ or
$\{p,\overline{p}\} \subseteq X$ for some variable $p$.
It is easy to see that if $S$ or $U$ is equivalent to a constant, then
determining whether $S \isoimpl U$ takes polynomial time, since
\begin{itemize}
\item If $S \equiv 1$, then $S \isoimpl U$ iff $U \equiv 1$.
\item If $S \equiv 0$ or $U \equiv 1$,
then $S \isoimpl U$.
\item If $U \equiv 0$, then $S \isoimpl U$ iff $S \equiv 0$.
\end{itemize}
It remains to consider the case that neither $S$ nor $U$ is equivalent to 
a constant.  We claim that in this case,
$S \isoimpl U$ iff the number of positive literals in $S$ is
greater or equal than the number of positive literals in $U$ and
the number of negative literals in $S$ is
greater or equal than the number of negative literals in $U$.
This completes the proof of Claim~\ref{clP}.
It remains to show the above claim.

First suppose that $\pi$ is a permutation of the variables
of $S \cup U$ such that $\pi(S) \Rightarrow U$.
Since $\pi(S) - \{1\}$ is a satisfiable set of literals,
it follows that, for all literals $\ell$, if $\pi(S) \Rightarrow \ell$,
then $\ell \in \pi(S)$.  This implies that $U - \{1\}
\subseteq \pi(S)$, and thus the number of positive literals in $\pi(S)$ is
greater or equal than the number of positive literals in $U$ and
the number of negative literals in $\pi(S)$ is
greater or equal than the number of negative literals in $U$.

For the converse, suppose that the number of positive literals in $S$ is
greater or equal than the number of positive literals in $U$ and
the number of negative literals in $S$ is
greater or equal than the number of negative literals in $U$.
Since no variable occurs both positively and
negatively in $S$ or $U$, it is easy to compute a permutation
$\pi$ of the variables in $S \cup U$
such that every variable that occurs positively
in $U$ is mapped to by a variable that occurs positively
in $S$ and such that every variable that occurs negatively
in $U$ is mapped to by a variable that occurs negatively
in $S$. It is immediate that $U - \{1\} \subseteq
\pi(S)$, and thus $S \isoimpl U$.
\end{proof}

\subsection{Lower bounds}

When proving dichotomy or dichotomy-like theorems for Boolean
constraints, the proofs of some of the lower bounds are
generally most involved.  In addition, proving lower bounds
for the case without constants is often a lot more involved than
the proofs for the case with constants.
This is particularly true in the case for isomorphism problems,
since here, we cannot introduce auxiliary variables.  

The approach
taken in~\cite{boe-hem-rei-vol:c:constraints,boe-hem-rei-vol:c:iso,boe-hem-rei-vol:t:con-iso-revised},
which examine the complexity of
the isomorphism problem for Boolean constraints, is to first
prove lower bounds for the case with constants, and then to show
that all the hardness reductions can be modified to obtain reductions for
the cases without constants.  

In contrast, in this paper we will prove the lower bounds directly
for the case without constants.  
We have chosen this approach since careful analysis of the cases
shows that proving the NP lower bounds boils down to proving 
NP-hardness for ten different cases (far fewer than in the
isomorphism paper). 

It should be noted that our NP lower bound results do not at all
follow from the lower bound results for the isomorphism problem.
This is also made clear by comparing Theorems~\ref{t:isotrich}
and~\ref{t:isoimp}: In some cases, the complexity jumps from
P to NP-complete, in other cases we
jump from GI-hard to NP-complete. 

\begin{lemma}
\label{l:implements}
Let $C$ be a $k$-ary constraint  such that
$C(x_1,\ldots, x_k)$  is not equivalent to
a conjunction of literals.
Then there exists a set of constraint applications of $C$ that
is equivalent to  one of the following ten constraint applications:
\begin{itemize}
\item
$t \wedge (x \vee y)$,
$\overline{f} \wedge t \wedge (x \vee y)$,
$\overline{f} \wedge  (\overline{x} \vee \overline{y})$,
$\overline{f} \wedge t \wedge (\overline{x} \vee \overline{y})$,
\item
$x \leftrightarrow y$,
$t \wedge (x \leftrightarrow y)$, 
$\overline{f} \wedge (x \leftrightarrow y)$,
$\overline{f} \wedge t \wedge (x \leftrightarrow y)$,
\item
$x \oplus y$, or
$\overline{f} \wedge t \wedge (x \oplus y)$.
\end{itemize}
\end{lemma}

\begin{proof}
Let $C$ be a $k$-ary constraint 
such that $C(x_1, \ldots, x_k)$
is not equivalent to a conjunction of literals.
First suppose that $C$ is not
2-affine.  It follows from~\cite[Lemma 24]{boe-hem-rei-vol:t:con-iso-revised}
that there exists a set $S$ of constraint applications of $C$
such that $S$ is equivalent to
$\overline{x} \wedge y$,
$\overline{x} \vee y$,
$x \oplus y$,
$x \leftrightarrow y$,
$t \wedge (\overline{x} \vee y)$, 
$t \wedge (x \leftrightarrow y)$, 
$t \wedge (x \vee y)$,
$\overline{f} \wedge (\overline{x} \vee y)$, 
$\overline{f} \wedge (x \leftrightarrow y)$,  or
$\overline{f} \wedge (\overline{x} \vee \overline{y})$.

If $S(x,y)$ is equivalent to $\overline{x} \vee y$, then
$S(x,y) \cup S(y,x)$ is equivalent to $x \leftrightarrow y$.
If $S(t,x,y)$ is equivalent to 
$t \wedge (\overline{x} \vee y)$,  then
$S(t,x,y) \cup S(t,y,x)$ is equivalent to $t \wedge (x \leftrightarrow y)$.
If $S(f,x,y)$ is equivalent to 
$\overline{f} \wedge (\overline{x} \vee y)$, then 
$S(f,x,y) \cup S(f,y,x)$ is equivalent to
$\overline{f} \wedge (x \leftrightarrow  y)$.

The only case that needs more work is the case that $S(x,y)$ is equivalent
to $\overline{x} \wedge y$.  From the proofs of
Theorems 15 and 17 of~\cite{boe-hem-rei-vol:t:con-iso-revised},
it follows that 
there exists a constraint application $A$ of $C$ such that
$A(0,1,x,y)$ is equivalent to $x \vee y$,
$\overline{x} \vee \overline{y}$, $\overline{x} \vee y$,
or $x \oplus y$.
It follows that $S(f,t) \cup \{A(f,t,x,y),A(f,t,y,x)\}$ is equivalent
to $\overline{f} \wedge t \wedge (x \vee y)$,
$\overline{f} \wedge t \wedge (\overline{x} \vee \overline{y})$, 
$\overline{f} \wedge t \wedge (x \leftrightarrow y)$, or
$\overline{f} \wedge t \wedge (x \oplus y)$.  This
completes the proof for the case that $C$ is not 2-affine.

To finish the proof of Lemma~\ref{l:implements},
suppose that $C$ is 2-affine.
Since $C(x_1, \ldots, x_k)$ is not equivalent to a conjunction
of literals, it is also not equivalent to 0, and
it follows from~\cite[Lemma 9]{boe-hem-rei-vol:c:iso} that
$C(x_1, \ldots, x_k)$ is equivalent to a formula of the form
\[\bigwedge_{x \in Z} \overline{x} \wedge
\bigwedge_{x \in O} x \wedge
\bigwedge_{i = 1}^\ell \left ( \left ( \bigwedge_{x \in X_i}x \wedge
\bigwedge_{y \in Y_i} \overline{y} \right ) \vee \left (
\bigwedge_{x \in X_i} \overline{x} \wedge
\bigwedge_{y \in Y_i} y  \right )\right ),\]
where $Z, O,  X_1, Y_1, \ldots, X_\ell,Y_\ell$ are pairwise
disjoint subsets of $\{x_1, \ldots, x_k\}$ such that
$X_i \cup Y_i \neq \emptyset$ for all $1 \leq i \leq \ell$.
Since $C(x_1, \ldots, x_k)$ is not equivalent to a conjunction
of literals,  there exists an $i$ such that $||X_i \cup Y_i ||
\geq 2$.

In $C(x_1, \ldots, x_k)$, replace all variables in $Z$ by $f$, and
all variables in $O$ by $t$. 

If for some $i$, $X_i \neq \emptyset$ and $Y_i \neq \emptyset$,
then replace all variables in $\bigcup_j X_j$ by $x$, and replace
all variables in $\bigcup_j Y_j$ by $y$.  In this case, the resulting
constraint application is equivalent to 
$x \oplus y$, $t \wedge (x \oplus y)$, $\overline{f} \wedge (x \oplus y)$, or
$\overline{f} \wedge t \wedge (x \oplus y)$.
In the second case, note that
$\{t \wedge (x \oplus y), t \wedge (t \oplus f)\}$
is a set of constraint applications of $C$ that is equivalent to
$\overline{f} \wedge t \wedge (x \oplus y)$.
In the third case, note that
$\{\overline{f} \wedge (x \oplus y), \overline{f} \wedge (t \oplus f)\}$
is a set of constraint applications of $C$ that is equivalent to
$\overline{f} \wedge t \wedge (x \oplus y)$.

If for all $i$, $X_i  = \emptyset$ or $Y_i = \emptyset$,
let $i$ be such that $||X_i|| \geq 2$ or $||Y_i || \geq 2$.
Replace one of the variables in $X_i \cup Y_i$ by $x$ and
replace all other variables in $\bigcup_j X_j \cup Y_j$ by $y$.
In this case, the resulting
constraint application is equivalent to 
$x \leftrightarrow y$, $t \wedge (x \leftrightarrow y)$,
$\overline{f} \wedge (x \leftrightarrow y)$, or
$\overline{f} \wedge t \wedge (x \leftrightarrow y)$.
\end{proof}

\subsection{The 10 reductions}

We will now show that in each
of the 10 cases of Lemma~\ref{l:implements},
the isomorphic implication problem is
NP-hard. Some work can be avoided by observing that
the isomorphic implication problem is computationally
equivalent to the same problem where every constraint
is replaced by a type of ``complement.''

In~\cite{hem:t:con-ph}, it is shown that the complexity
of (quantified) satisfiability problems for a set
of constraints ${\cal C}$ is the same as the complexity
of the same problem for the set of constraints
${\cal C}^c$, where ${\cal C}^c$ is defined as follows.

\begin{definition}[\cite{hem:t:con-ph}]
\begin{enumerate}
\item
For $C$ a $k$-ary constraint,
$C^c$ is the $k$-ary constraint such that for all $s \in \{0,1\}^k$,
$C^c(s) = C(\overline{s})$, where, as in the definition
of complementive, $\overline{s} = (1 - s_1)(1-s_2) \cdots (1-s_k)$ for $s =
s_1s_2 \cdots s_k$.
\item For ${\cal C}$ a finite set of constraints,
${\cal C}^c = \{C^c \ | \ C \in {\cal C}\}$.
\item For $S$ a set of constraint applications of ${\cal C}$,
$S^c = \{C^c(z_1, \ldots, z_k) \ | \
C(z_1, \ldots, z_k) \in S\}$.
\end{enumerate}
\end{definition}

It is easy to see that any isomorphism from $S$ to $U$ is also
an isomorphism from $S^c \cong U^c$ (and vice versa). This
implies the following.

\begin{lemma}
\label{l:complement}
$\isoimp({\cal C})$ is computationally equivalent to $\isoimp({\cal C}^c)$.
\end{lemma}

As mentioned in the introduction and proven in Section~\ref{s:analog},
the NP-complete subgraph isomorphism problem 
is closely related to the isomorphic implication problem
for sets of constraint applications of $\lambda xy. x \vee y$, in
the following way:
For a graph $\widehat{G}$, let $S(\widehat{G})$ be defined
as $\{x_i \vee x_j \ | \ \{i,j\} \in E(\widehat{G})\}$.
It is easy to see that for two graphs $G$ and $H$ without isolated
vertices, $G$ contains a subgraph isomorphic to $H$ if and only
if $S(G)$ isomorphically implies $S(H)$. This correspondence is also
the reason for the GI-hardness for the isomorphism problem
for sets of constraint applications of
$\lambda xy. x \vee y$~\cite{bor-ran-ste:j:boolean-equivalence,boe-hem-rei-vol:c:constraints}.

We will use the observation above to prove NP-hardness
for constraints that are similar to $\lambda xy. x \vee y$, namely,
we will reduce the subgraph isomorphism problem to
the isomorphic implication problems for
$\lambda txy . t \wedge (x \vee y)$, 
$\lambda ftxy . \overline{f} \wedge t \wedge (x \vee y)$, 
$\lambda fxy. \overline{f} \wedge (\overline{x} \vee \overline{y})$, and
$\lambda ftxy . \overline{f} \wedge t \wedge 
(\overline{x} \vee \overline{y})$. 

\begin{ourclaim}
\label{cl:or}
\begin{enumerate}
\item
$\isoimp(\{\lambda txy. t \wedge (x \vee y)\})$
is \np-hard.
\item
$\isoimp(\{\lambda ftxy. \overline{f} \wedge t \wedge (x \vee y)\})$
is \np-hard.
\item
$\isoimp(\{\lambda fxy. \overline{f} \wedge (\overline{x} \vee
\overline{y})\})$ 
is \np-hard.
\item
$\isoimp(\{\lambda ftxy. \overline{f} \wedge t \wedge (\overline{x} \vee
\overline{y})\})$ 
is \np-hard.
\end{enumerate}
\end{ourclaim}

\begin{proof}
\begin{enumerate}
\item
Let $G$ and $H$ be two graphs without isolated vertices.
For $\widehat{G}$ a graph, define
\[S(\widehat{G}) = \{t \wedge (x_i \vee x_j) \ | \ \{i,j\} \in
E(\widehat{G})\}.\]

We claim that $G$ contains a subgraph isomorphic to $H$ if and only if
$S(G) \isoimpl S(H)$.

First suppose that $G'$ is a subgraph of $G$ and that $G'$ is isomorphic
to $H$. Then $S(G') \subseteq S(G)$ and
there exists a bijection $\pi$ from $V(G')$ to $V(H)$ such that
$\pi(G') = H$, which implies that $S(\pi(G')) = S(H)$.
Let $\rho$ be a permutation on the set
$\{t\} \cup \{x_i \ | \ i \in V(G) \cup V(H)\}$ such that 
$\rho(t) = t$ and $\rho(x_i) = x_{\pi(i)}$ for
all $i \in V(G')$.  It is immediate that
$\rho(S(G')) = S(\pi(G')) = S(H)$ and that
$\rho(S(G')) \subseteq \rho(S(G))$. It follows that
$S(H) \subseteq \rho(S(G))$, and thus
$S(G) \isoimpl S(H)$.

For the converse, suppose that there exists a permutation
$\rho$ on the variables occurring in $S(G) \cup S(H)$
such that $\rho(S(G)) \Rightarrow S(H)$. First note that
such a $\rho$ must map $t$ to $t$, since, for any graph $\widehat{G}$ without
isolated vertices, $t$ is the unique variable $z$ such that
$S(\widehat{G}) \Rightarrow z$.   Also note that for all
graphs $\widehat{G}$, if $S(\widehat{G}) \Rightarrow t \wedge (x_i \vee x_j)$,
then $t \wedge (x_i \vee x_j) \in S(\widehat{G})$.

It is easy to see that if
$\rho(S(\widehat{G})) \Rightarrow t \wedge (x_i \vee x_j)$,
then $t \wedge (x_i \vee x_j) \in \rho(S(\widehat{G}))$.
(For if it were not, setting $t$ to true, $x_i$ and $x_j$ to false,
and all other $x$-variables to true would satisfy $\rho(S(\widehat{G}))$.)
It follows that $S(H) \subseteq \rho(S(G))$.
Let $G'$ be the graph isomorphic to $H$ such that
$S(H) = \rho(S(G'))$.  Then $\rho(S(G')) \subseteq \rho(S(G))$,
and thus $G'$ is a subgraph of $G$.

\item For $\widehat{G}$ a graph, define
\[S'(\widehat{G}) = \{\overline{f} \wedge t \wedge (x_i \vee x_j)
\ | \ \{i,j\} \in E(\widehat{G})\}.\]

We claim that for any graphs $G$ and $H$ without isolated vertices,
$S(G) \isoimpl S(H)$ if and only if $S'(G) \isoimpl S'(H)$.

First suppose that $\rho$ is a permutation of the variables
occurring in $S(G) \cup S(H)$ such that $\rho(S(G)) \Rightarrow S(H)$.
If we extend $\rho$ by letting $\rho(f) = f$, then
$\rho(S'(G)) \Rightarrow S'(H)$.  For the converse, suppose that 
$\rho$ is a permutation of the variables
occurring in $S'(G) \cup S'(H)$ such that $\rho(S'(G)) \Rightarrow S'(H)$.
Then $\rho(f) = f$, since,  for any graph $\widehat{G}$ without
isolated vertices, $f$ is the unique variable $z$ such that
$S(\widehat{G}) \Rightarrow \overline{z}$.  Since,
for any graph $\widehat{G}$,  $S'(\widehat{G})$ is
equivalent to $\overline{f} \wedge S(\widehat{G})$
and $f$ does not occur in $S(\widehat{G})$, it follows
immediately that $\rho(S(G)) \Rightarrow S(H)$.

\item
Note that $(\lambda fxy . \overline{f} \wedge (\overline{x}
\vee \overline{y}))^c =
\lambda fxy. f \wedge  (x \vee y)$.
The result follows immediately from part 1 of this claim
and Lemma~\ref{l:complement}.

\item
Note that $(\lambda ftxy . \overline{f} \wedge t \wedge (\overline{x}
\vee \overline{y}))^c =
\lambda ftxy . f \wedge \overline{t} \wedge (x \vee y)$.
The result follows immediately from part 2 of this claim
and Lemma~\ref{l:complement}.
\end{enumerate}
\end{proof}

The remaining 6 constraints behave differently.  In these
cases, the isomorphism problem is in P. Thus, GI does not reduce
to these isomorphism problems (unless GI is in P), 
and there does not seem to be a simple reduction
from the subgraph isomorphism problem to
the isomorphic implication problem.
In these cases, we will prove NP-hardness by reduction from a suitable
partitioning problem, namely, the unary version
of the problem $\threepart$~\cite[Problem SP15]{gar-joh:b:int}.

\begin{definition}\cite{gar-joh:b:int}
$\unarythreepart$ is the problem, given 
a set $A$ of $3m$ elements, $B \in \Z^+$ a bound (in unary),
and for each $a\in A$, a size $s(a) \in \Z^+$ (in unary)
such that $B/4 < s(a) < B/2$ and such that
$\sum_{a\in A} s(a) = mB$, to decide
whether $A$ can be partitioned into 
$m$ disjoint sets $A_1,\ldots,A_m$ such that $\sum_{a\in A_i}s(a)=B$ for 
$1 \leq i \leq m$.
\end{definition}

\begin{theorem}[\cite{gar-joh:b:int}]
$\unarythreepart$ is \np-complete.
\end{theorem}

\begin{ourclaim}
\label{cl:iff}
\begin{enumerate}
\item
$\isoimp(\{\lambda xy. x \leftrightarrow y \})$ is \np-hard.
\item
$\isoimp(\{\lambda txy. t \wedge (x \leftrightarrow y) \})$ is
\np-hard.
\item
$\isoimp(\{\lambda fxy. \overline{f} \wedge (x \leftrightarrow y)\})$
is \np-hard.
\item
$\isoimp(\{\lambda ftxy. \overline{f} \wedge t \wedge (x \leftrightarrow y)\})$
is \np-hard.
\end{enumerate}
\end{ourclaim}

\begin{proof}
\begin{enumerate}
\item
Let $A$ be a set with $3m$ elements, $B \in \Z^+$ a bound (in unary),
and for each $a\in A$, let  $s(a) \in \Z^+$ be a size (in unary)
such that $\sum_{a\in A} s(a) = mB$.

Let $X_1, \ldots, X_m$ be $m$ pairwise disjoint sets
of variables, each of size $B$.  Let
\[S = \{ x \leftrightarrow x' \ | \ x, x' \in X_i  \mbox{ for some } 
i\}.\]

Let $\{\widehat{X}_a \ | \ a \in A\}$ be a collection
of $3m$ pairwise disjoint sets of variables such that
$||\widehat{X}_a|| = s(a)$ for all $a \in A$,
and such that
\[\bigcup_{a \in A} \widehat{X}_a = \bigcup_{i = 1}^m X_i.\]

Let
\[U = \{ x \leftrightarrow x' \ | \ x,x' \in \widehat{X}_a
\mbox{ for some } a \in A\}.\]

Note that since $B$ and the $s(a)$'s are given in unary, $S$ and $U$ can be
computed in polynomial time.

We claim that $A$ can be partitioned into 
$m$ disjoint sets $A_1,\ldots,A_m$ such that $\sum_{a\in A_i}s(a)=B$ for 
$1 \leq i \leq m$ if and only if $S \isoimpl U$.

First suppose that
$A_1,\ldots,A_m$ is a partition of $A$ such that $\sum_{a\in A_i}s(a)=B$ for 
$1 \leq i \leq m$. Define a permutation $\pi$ on 
$\bigcup_{i=1}^m X_i$ such that
for all $i$,  $\pi(X_i) = \bigcup_{a \in A_i}\widehat{X}_a$.
Let $(x \leftrightarrow x') \in U$. Then, for some $a \in A$,
$x,x' \in \widehat{X}_a$. Then 
there exists an $i$ such that $\pi^{-1}(x)$ and $\pi^{-1}(x')$
are elements of $X_i$, which implies that 
$(\pi^{-1}(x) \leftrightarrow \pi^{-1}(x')) \in S$, and thus
$(x \leftrightarrow x') \in \pi(S)$. 
It follows that $U \subseteq \pi(S)$, and thus
$\pi(S) \Rightarrow U$.

For the converse, suppose $\pi$ is a permutation of
$\bigcup_{i=1}^m X_i $ such that
$\pi(S) \Rightarrow U$.
Let
$A_i = \{a \ | \  \pi(X_i) \cap \widehat{X}_a \neq \emptyset\}$.
We claim that $A_1, \ldots, A_{m}$ is a desired partition.

By definition, it is immediate that $\bigcup_{i=1}^m A_i = A$.
Next suppose that $A_i \cap A_j \neq \emptyset$, for some $i \neq j$.  Then
for some $z \in X_i$, $z' \in X_j$, there exists an
$a \in A$ such that $\pi(z), \pi(z') \in \widehat{X}_a$.
Then $(\pi(z) \leftrightarrow \pi(z')) \in U$.
But it easy to see that there exists a satisfying assignment of $S$
such that $z$ is true and $z'$ is false.  Thus, 
$S \not \Rightarrow (z \leftrightarrow z')$.  This implies that
$\pi(S) \not \Rightarrow (\pi(z) \leftrightarrow \pi(z'))$.
But this contradicts the fact that $\pi(S) \Rightarrow U$.

It follows that $A_i \cap A_j = \emptyset$ for all $i \neq j$ and
it follows that
$\pi(X_i) = \bigcup_{a \in A_i}\widehat{X}_a$.
Since $\pi$ is an injection,
$||X_i|| = ||\bigcup_{a \in A_i}\widehat{X}_a||$, and since
the $\widehat{X}_a$'s are pairwise disjoint,
$||\bigcup_{a \in A_i}\widehat{X}_a|| = 
\Sigma_{a \in A_i}s(a)$.  Since $||X_i|| = B$,
it follows that $\Sigma_{a \in A_i}s(a) = B.$
This completes the NP-hardness proof for 
$\isoimp(\{\lambda xy. x \leftrightarrow y\})$.

\item
Now consider the case where the constraint
is $\lambda txy. t \wedge (x \leftrightarrow y)$.
Let

\[S' = \{t \wedge
(x \leftrightarrow x') \ | \ x,x' \in X_i \mbox{ for some } i\}\]

and
\[U' = \{t \wedge  (x \leftrightarrow x') \ |
\ x,x' \in \widehat{X}_a \mbox{ for some } a \in A\}.\]

We claim that  $S' \isoimpl U'$ if and only if $S \isoimpl U$.
The left-to-right direction is immediate: Simply extend the
permutation $\pi$ on the variables occurring in $S \cup U$
such that $\pi(S) \Rightarrow U$ by
letting  $\pi(t) = t$.  Then
$\pi(S') \Rightarrow \pi(U')$.

For the converse, note that $S' \equiv t \wedge S$
and $U' \equiv t \wedge U$.
Let $\pi$ be such that $\pi(S') \Rightarrow U'$. Note that
$t$ is the unique variable $z$ such that $S' \Rightarrow z$.
It follows that $\pi$ maps $t$ to $t$. 
Since $t$ does not occur in $S$ and $U$, it follows
that $\pi(S) \Rightarrow U$.

\item
$(\lambda fxy. \overline{f} \wedge (x \leftrightarrow y))^c$
is equivalent to $\lambda txy. t \wedge (x \leftrightarrow y)$.
The result follows immediately
from part 2 of this claim and Lemma~\ref{l:complement}.

\item
Now consider the case where the constraint
is $\lambda ftxy. \overline{f} \wedge t \wedge (x \leftrightarrow y)$.
Let

\[S'' = \{\overline{f} \wedge t \wedge
(x \leftrightarrow x') \ | \ x,x' \in X_i \mbox{ for some } i\}\]

and
\[U'' = \{\overline{f} \wedge t \wedge  (x \leftrightarrow x') \ |
\ x,x' \in \widehat{X}_a \mbox{ for some } a \in A\}.\]

We claim that $S'' \isoimpl U''$ if and only if $S \isoimpl U$.
The left-to-right direction is immediate: Simply extend the
permutation $\pi$ on the variables occurring in $S \cup U$
such that $\pi(S) \Rightarrow U$ by
letting $\pi(f) = f$ and $\pi(t) = t$.  Then
$\pi(S'') \Rightarrow \pi(U'')$.

For the converse, note that $S'' \equiv \overline{f} \wedge t \wedge S$
and $U'' \equiv \overline{f} \wedge t \wedge U$.
Let $\pi$ be such that $\pi(S'') \Rightarrow U''$. Note that $f$ is the
unique variable $z$ such that $S'' \Rightarrow \overline{z}$ and
and that $t$ is the unique variable $z$ such that $S'' \Rightarrow z$.
It follows that $\pi$ maps $t$ to $t$ and $f$ to $f$.
Since $t$ and $f$ do not occur in $S$ and $U$, it follows
that $\pi(S) \Rightarrow U$.

\end{enumerate}
\end{proof}

For the final two cases, we adapt the proof from the previous claim.

\begin{ourclaim}
\label{cl:xor}
\begin{enumerate}
\item
$\isoimp(\{\lambda xy. x \oplus y \})$ is \np-hard.
\item
$\isoimp(\{\lambda ftxy. \overline{f} \wedge t \wedge (x \oplus y)\})$
is \np-hard.
\end{enumerate}
\end{ourclaim}

\begin{proof}
\begin{enumerate}
\item
Let $A$ be a set with $3m$ elements, $B \in \Z^+$ a bound (in unary),
and for each $a\in A$, let $s(a) \in \Z^+$ be a size (in unary)
such that $\sum_{a\in A} s(a) = mB$.

Let $X_1, \ldots, X_m, Y_1, \ldots, Y_m$ be $2m$ pairwise disjoint sets
of variables, each  of size $B$.  Let
\[S = \{ x \oplus y \ | \ x \in X_i \mbox{ and } y \in Y_i \mbox{ for some } 
i\}.\]

Let $\{\widehat{X}_a, \widehat{Y}_a \ | \  a \in A\}$ be a collection
of $6m$ pairwise disjoint sets of variables such that
$||\widehat{X}_a|| = ||\widehat{Y}_a|| = s(a)$ for all $a \in A$,
and such that
\[\bigcup_{a \in A} (\widehat{X}_a  \cup \widehat{Y}_a) = 
\bigcup_{i = 1}^m (X_i \cup Y_i).\]

Let
\[U = \{ x \oplus y \ | \ x \in \widehat{X}_a
\mbox{ and } y \in \widehat{Y}_a \mbox{ for some } a \in A\}.\]

Note that since $B$ and the $s(a)$'s are given in unary, $S$ and $U$ can be
computed in polynomial time.

We claim that $A$ can be partitioned into 
$m$ disjoint sets $A_1,\ldots,A_m$ such that $\sum_{a\in A_i}s(a)=B$ for 
$1 \leq i \leq m$ if and only if $S \isoimpl U$.

First suppose that
$A_1,\ldots,A_m$ is a partition of $A$ such that $\sum_{a\in A_i}s(a)=B$ for 
$1 \leq i \leq m$. Define a permutation $\pi$ on 
$\bigcup_{i=1}^m(X_i \cup Y_i)$ such that
for all $i$,  $\pi(X_i) = \bigcup_{a \in A_i}\widehat{X}_a$ and
$\pi(Y_i) = \bigcup_{a \in A_i}\widehat{Y}_a$.
Consider an arbitrary element of $U$, say $x \oplus y$ for 
$x \in \widehat{X}_a$ and $y \in \widehat{Y}_a$.
Then there exists an $i$ such that 
$\pi^{-1}(x) \in X_i$ and $\pi^{-1}(y) \in Y_i$ (or vice versa),
which implies that
$(\pi^{-1}(x) \oplus \pi^{-1}(y)) \in S$, and thus
$(x \oplus y) \in \pi(S)$.
It follows that $U \subseteq \pi(S)$, and thus
$\pi(S) \Rightarrow U$.

For the converse, suppose $\pi$ is a permutation of
$\bigcup_{i=1}^m(X_i \cup Y_i)$ such that
$\pi(S) \Rightarrow U$.

Let \[A_i = \{a \ | \  \pi(X_i \cup Y_i) \cap (\widehat{X}_a \cup
\widehat{Y}_a) \neq \emptyset\}.\]

We claim that $A_1, \ldots, A_{m}$ is the desired partition. 

By definition, it is immediate that $\bigcup_{i=1}^m A_i = A$.
Next suppose that $A_i \cap A_j \neq \emptyset$, for some $i \neq j$.  Then
for some $z \in X_i \cup Y_i$, $z' \in X_j \cup Y_j$, there exists an
$a \in A$ such that $\pi(z), \pi(z') \in \widehat{X}_a \cap \widehat{Y}_a$.

Then $U \Rightarrow (\pi(z) \oplus \pi(z'))$ or $U \Rightarrow (\pi(z)
\leftrightarrow \pi(z'))$.
But it easy to see that there exists a satisfying assignment of $S$
such that $z$ is true and $z'$ is false, and
that there exists a satisfying assignment of $S$
such that $z$ is true and $z'$ is true.
Thus, $S \not \Rightarrow (z \oplus z')$ and
$S \not \Rightarrow (z \leftrightarrow z')$.
This implies that
$\pi(S) \not \Rightarrow (\pi(z) \oplus \pi(z'))$ and
$\pi(S) \not \Rightarrow (\pi(z) \leftrightarrow \pi(z'))$.
But this contradicts the fact that $\pi(S) \Rightarrow U$.

It follows that $A_i \cap A_j = \emptyset$ for all $i \neq j$ and
it follows that
$\pi(X_i \cup Y_i) = \bigcup_{a \in A_i}(\widehat{X}_a \cup \widehat{Y}_a)$.
Since $\pi$ is an injection,
$||X_i \cup Y_i|| = ||\bigcup_{a \in A_i}(\widehat{X}_a \cup \widehat{Y}_a)||
= \Sigma_{a \in A_i}2s(a)$.  Since $||X_i \cup Y_i|| = 2B$
it follows that $\Sigma_{a \in A_i}2s(a) = 2B$, which, of course,
implies that $\Sigma_{a \in A_i}s(a) = B$, as required.

\item
Now consider the case where the constraint is $\lambda ftxy.\overline{f}
\wedge t \wedge (x \oplus y)$. Let

\[S' = \{\overline{f} \wedge t \wedge
x \oplus y \ | \ x \in X_i \mbox{ and } y \in Y_i \mbox{ for some } i\}\]

and
\[U' = \{\overline{f} \wedge t \wedge x \oplus y \ | \ x \in \widehat{X}_a
\mbox{ and } y \in \widehat{Y}_a \mbox{ for some } a \in A\}.\]

We claim that $S' \isoimpl U'$ if and only if $S \isoimpl U$.
The left-to-right direction is immediate: Simply extend the
permutation $\pi$ on the variables occurring in $S \cup U$
such that $\pi(S) \Rightarrow U$ by
letting $\pi(f) = f$ and $\pi(t) = t$.  Then
$\pi(S') \Rightarrow \pi(U')$.

For the converse, note that $S' \equiv \overline{f} \wedge t \wedge S$
and $U' \equiv \overline{f} \wedge t \wedge U$.
Let $\pi$ be such that $\pi(S') \Rightarrow U'$. Note that $f$ is the
unique variable $z$ such that $S' \Rightarrow \overline{z}$ and
and that $t$ is the unique variable $z$ such that $S' \Rightarrow z$.
It follows that $\pi$ maps $t$ to $t$ and $f$ to $f$.
Since $t$ and $f$ do not occur in $S$ and $U$, it follows
that $\pi(S) \Rightarrow U$.

\end{enumerate}
\end{proof}

To complete the proof of Theorem~\ref{t:isoimp}, it
remains to show the following claim.

\begin{ourclaim}
\label{cl:conp}
Let \cset\ be a finite set of constraints. If \cset\ is not Schaefer, then 
$\isoimp(\cset)$ is \conp-hard.
\end{ourclaim}

\begin{proof}
The exact same reductions that show the coNP-hardness
for $\iso({\cal C})$ from~\cite[Claim 19]{boe-hem-rei-vol:c:constraints}
also show coNP-hardness for $\isoimp({\cal C})$.  This is
because for all pairs of sets of constraint applications $(S,U)$
of ${\cal C}$ that these reductions map to, it holds that $U \isoimpl S$.
Under this condition, $S \cong U$ if and only if $S \isoimpl U$.
\end{proof}

\section{Toward a Trichotomy Theorem}
\label{s:conjecture}

The current main theorem (Theorem~\ref{t:isoimp}) is not a trichotomy
theorem, since for ${\cal C}$ not Schaefer, it states that
$\isoimp({\cal C})$ is NP-hard, coNP-hard, and in $\parallelnp$. The large gap
between the lower and upper bounds is not very satisfying.
We conjecture that the current lower bounds for $\isoimp({\cal C})$
for ${\cal C}$ not 
Schaefer can be raised to $\parallelnp$ lower bounds, which
would give the following trichotomy theorem.

\begin{ourconjecture}
\label{c:isoimp}
Let \cset\ be a finite set of constraints.
\begin{enumerate}
\item If every constraint in \cset\ is equivalent to a constant or
a conjunction of literals,
then $\isoimp(\cset)$ and $\isoimp_c(\cset)$ are in \p.
\item Otherwise, if \cset\ is Schaefer, then $\isoimp(\cset)$ and
$\isoimp_c(\cset)$ are \np-complete.
\item If \cset\ is not Schaefer, then $\isoimp(\cset)$ and 
$\isoimp_c(\cset)$ are $\parallelnp$-complete.
\end{enumerate}
\end{ourconjecture}

We believe this conjecture for two reasons.  First of all,
it is quite common for problems that are NP-hard, coNP-hard, and
in $\parallelnp$ to end up being $\parallelnp$-complete.
(For an overview of this phenomenon, 
see~\cite{hem-hem-rot:j:raising-lower-bounds}.)
Secondly, we will prove $\parallelnp$ lower bounds for some cases
in Theorem~\ref{t:parallelnphard}.

To raise NP and coNP lower bounds to $\parallelnp$ lower bounds, the
following theorem by Wagner often plays a crucial role, which it
will also do in our case.

\begin{theorem}[\cite{wag:j:more-on-bh}]
\label{t:wagner}
Let $L$ be a language.
If there exists a polynomial-time computable function $h$ such that
\[||\{i \ | \ \phi_i \in \sat\}|| \mbox{ is odd iff }
h(\phi_1, \ldots, \phi_{2k} ) \in L\]
for all $k \geq 1$ and all Boolean formulas  $\phi_1, \ldots, \phi_{2k}$
such that $\phi_{i} \in \sat \Rightarrow \phi_{i+1} \in \sat$,
then $L$ is $\parallelnp$-hard.
\end{theorem}

The basic idea behind applying Wagner's theorem to turn an NP
lower bound and a coNP lower bound into a $\parallelnp$ lower
bound is the following. 

\begin{lemma}
\label{l:wagner}
Let $L$ be a language.  If $L$ is \np-hard and
\conp-hard, and ($L$ has polynomial-time computable
and- and $\omega$-or functions or
$L$ has polynomial-time computable or- and $\omega$-and functions), then
$L$ is $\parallelnp$-hard.\footnote{
An or-function for a language $L$ is a function $f$ such that for
all $x,y \in \Sigma^*$, $f(x,y) \in L$ iff $x \in L$ or $y \in L$.
An $\omega$-or-function for a language $L$ is a function $f$ such that for
all $x_1, \ldots, x_n \in \Sigma^*$, $f(x_1, \ldots, x_n) \in L$
iff $x_i \in L$ for some $i$; and-functions are defined
similarly~\cite{koe-sch-tor:b:graph-iso}.}
\end{lemma}

\begin{proof}
First suppose that $L$ has a polynomial-time computable
and-function {\em and}, and a polynomial-time computable
$\omega$-or function {\em or}. Let $f$ be a reduction
from $\overline{\sat}$ to $L$ and let $g$ be
a reduction from $\sat$ to $L$.

Let $k \geq 1$ and let $\phi_1, \ldots, \phi_{2k}$ be formulas
such that $\phi_{i} \in \sat \Rightarrow \phi_{i+1} \in \sat$.
Note that $||\{i \ | \ \phi_i \in \sat\}||$  is odd if and only if
there exists an $i$ such that $1 \leq i \leq k$, 
$\phi_{2i-1} \not \in \sat$, and  $\phi_{2i} \in \sat$.

Define $h(\phi_1, \ldots, \phi_{2k})$ as
\[\mbox{\em or} \left ( \mbox{\em and}(f(\phi_1),g(\phi_2)),
\mbox{\em and}(f(\phi_3),g(\phi_4)), \ldots,
\mbox{\em and}(f(\phi_{2k-1}),g(\phi_{2k})) \right ). \]

It is immediate that $h$ is computable in polynomial-time
and there exists an $i$ such that $1 \leq i \leq k$, 
$\phi_{2i-1} \not \in \sat$, and  $\phi_{2i} \in \sat$ if and only if
$h(\phi_1, \ldots, \phi_{2k}) \in L$.  It follows that
$L$ is $\parallelnp$-hard by Theorem~\ref{t:wagner}.

Now consider the case  that $L$ has a polynomial-time computable
or-function, and a polynomial-time computable $\omega$-and function.
Then $\overline{L}$ has a polynomial-time computable
and-function, and a polynomial-time computable $\omega$-or function.
By the argument above, $\overline{L}$ is $\parallelnp$-hard.
Since $\parallelnp$ is closed under complement, it follows that
$L$ is $\parallelnp$-hard.
\end{proof}

Agrawal and Thierauf~\cite{agr-thi:j:boolean-isomorphism}
proved that the Boolean isomorphism problem has
$\omega$-and and $\omega$-or functions.  Since the Boolean isomorphism
problem is trivially coNP-hard, we obtain the following corollary.

\begin{corollary}
If the Boolean isomorphism problem is \np-hard, then it
is $\parallelnp$-hard.
\end{corollary}

Unfortunately, Agrawal and Thierauf's $\omega$-or function does not work
for Boolean isomorphic implication.  Their $\omega$-and function
seems to work for Boolean isomorphic implication, but since
this function or's two formulas together, it will not work
for sets of constraint applications.

To prove our $\parallelnp$ lower bounds,
we need to come up with completely new constructions.
In the proof, we will use the following lemma.

\begin{lemma}
\label{l:variables}

Let $S$ and $U$ be two sets of constraint applications, let $X$ be 
the set of variables occurring in $S$, and let $Y$ be the set of
variables occurring in $U$. If $S \isoimpl U$, $||X|| \geq ||Y||$, and
$X \cap Y = \emptyset$, then there exists a permutation $\pi$ of $X \cup Y$
such that $\pi(S) \Rightarrow U$ and $\pi(Y) \cap Y = \emptyset$.
\end{lemma}

\begin{proof}
Let $\pi'$ be a permutation such that $\pi'(S) \Rightarrow U$. If
there exist $y, y' \in Y$ with $\pi'(y') = y$, then, since 
$||X|| \geq ||Y||$, 
there exist $x,x' \in X$ such that $\pi'(x')=x$. Construct
a new permutation $\rho$ as follows: $\rho(y')=x$, $\rho(x')=y$, 
$\rho(z) = \pi'(z)$ for all $z \in (X \cup Y) - \{x',y'\}$. We will show
that $\rho(S) \Rightarrow U$. By repeatedly applying this construction,
we get a permutation $\pi$ such that $\pi(S) \Rightarrow U$ and
$\pi(Y) \subseteq X$. Since
$X \cap Y = \emptyset$, it follows that $\pi(Y) \cap Y = \emptyset$.

It remains to show that $\rho(S) \Rightarrow U$. For this,
suppose that $\rho(S) \not \Rightarrow U$. Let $Z$ be a list of the
variables in $(S \cup U)-\{x',y'\}$. Then 
$\rho(S)(Z,x',y') \not \Rightarrow U(Z,x',y')$ and thus there exist 
$a,b \in \{0,1\}$ and $s \in \{0,1\}^{||Z||}$, such that $\rho(S)(s,a,b) = 1$
and $U(s,a,b)=0$. Since $x'$ does not occur in $U$ and $y'$ does not
occur in $S$, we also have $\rho(S)(s,a,\overline{b}) = 1$
and $U(s,\overline{a},b)=0$. Since $\pi'(S)(Z,x,y) \Rightarrow U$ it follows
that $\pi'(S)(s,\overline{a},b) = \pi'(S)(s,a,b)=0$. Because $y'$ does not
occur in $S$, also $\pi'(S)(s,\overline{a}, \overline{b}) = 
\pi'(S)(s, a, \overline{b}) = 0$. So, $\pi'(S)(s,x,y) \equiv 0$, but
$\pi'(S)(s,b,a)=\rho(S)(s,a,b)=1$.
\end{proof}

\begin{theorem}
\label{t:parallelnphard}
Let ${\cal D}$ be a set of constraints that is 0-valid, 
1-valid, not complementive, and not Schaefer. Let ${\cal C} = {\cal D} 
\cup \{\lambda xy. x \vee y\}$.
Then $\isoimp({\cal C})$ is $\parallelnp$-complete.
\end{theorem}

\begin{proof}
By Theorem~\ref{t:isoimp}, $\isoimp({\cal C})$ is in $\parallelnp$.
Thus it suffices to show that $\isoimp({\cal C})$ is $\parallelnp$-hard.
Let $k \geq 1$ and let $\phi_1, \ldots, \phi_{2k}$ be formulas
such that $\phi_{i} \in \sat \Rightarrow \phi_{i+1} \in \sat$.
We will construct a polynomial-time computable function $h$ such that
\[||\{i \ | \ \phi_i \in \sat\}|| \mbox{ is odd iff }
h(\phi_1, \ldots, \phi_{2k} ) \in \isoimp({\cal C}).\]
By Theorem~\ref{t:wagner}, this proves that $\isoimp({\cal C})$ is
$\parallelnp$-hard.

Note that $||\{i \ | \ \phi_i \in \sat\}||$  is odd if and only if
there exists an $i$ such that $1 \leq i \leq k$, 
$\phi_{2i-1} \not \in \sat$, and  $\phi_{2i} \in \sat$.
This is a useful way of looking at it, and we will
prove that there exists an $i$ such that $1 \leq i \leq k$, 
$\phi_{2i-1} \not \in \sat$ and  $\phi_{2i} \in \sat$ if and
only if $h(\phi_1, \ldots, \phi_{2k} ) \in \isoimp({\cal C}).$

{From} Theorem~\ref{t:isoimp} we know that $\isoimp({\cal C})$ is NP-hard
and coNP-hard,
and thus there exist (polynomial-time many-one) reductions  from
\sat\ to  $\isoimp({\cal C})$ and from $\overline{\sat}$ to
$\isoimp({\cal C})$.  We will follow the idea of the proof
of Lemma~\ref{l:wagner}, but we 
will look in more detail at the reductions, so that we
can restrict the sets of constraint applications that we have to
handle.

Let $f$ be a polynomial-time computable function such that for all $\phi$,
$f(\phi)$ is  a set of constraint applications of ${\cal D}$ and
\[\phi \in \overline{\sat} \mbox { iff } f(\phi) \isoimpl
\bigcup_{1 \leq j, \ell \leq n}  \{x_{j} \rightarrow x_{\ell}\}
.\]
Here $x_1, \ldots, x_n$ are exactly all variables in $f(\phi)$.
Such a function exists, since  $\overline{\sat}$ is reducible
to $\overline{\csp_{\neq \vec{0}, \vec{1}}({\cal D})}$
($\csp_{\neq \vec{0}, \vec{1}}({\cal D})$ is the problem
of deciding whether a set of constraint applications of ${\cal D}$
has a satisfying assignment other than $\vec{0}$ and $\vec{1}$),
which is reducible
to $\isoimp({\cal D})$ via a reduction that satisfies the properties above.
(See the proofs of Claim~\ref{cl:conp}
and~\cite[Claims 19 and 14]{boe-hem-rei-vol:c:constraints}.)

Let $g$ be a polynomial-time computable function such that for all $\phi$,
$g(\phi)$ is  a set of constraint applications of $\lambda xy. x \vee y$
without duplicates (i.e., if $z \vee z' \in g(\phi)$, then $z \neq z'$) and

\[\phi \in \sat \mbox { iff } g(\phi) \isoimpl \{y_j \vee y_{j+1} \ | \
1 \leq j < n\}.\]
Here $y_1, \ldots, y_n$ are exactly all variables occurring in $g(\phi)$.
Such a function exists, since  $\sat$ is reducible to
HAMILTONIAN PATH, which is reducible to 
$\isoimp(\{\lambda xy.x \vee y\})$ via a reduction that satisfies the
properties above.  (Basically, use the standard translation from
graphs to sets of constraint applications of $\lambda xy . x \vee y$:
For $G$ a connected graph on vertices $\{1, \ldots, n\}$,
let $g(G) = \{y_i \vee y_j \ | \ \{i,j\} \in E(G)\}$.)

Recall that we need to construct
a polynomial-time computable function $h$ with the property that
there exists an $i$ such that $1 \leq i \leq k$, 
$\phi_{2i-1} \not \in \sat$, and $\phi_{2i} \in \sat$
if and only if $h(\phi_1, \ldots, \phi_{2k} ) \in \isoimp({\cal C})$.

In order to construct $h$, we will apply the coNP-hardness reduction
$f$ on $\phi_i$ for odd $i$, and the NP-hardness reduction $g$ on
$\phi_i$ for even $i$.  It will be important to make sure that all obtained
sets of constraint applications are over disjoint sets of variables.

For every $i$, $1 \leq i \leq k$, we define $O_i$ to be the set
of constraint applications $f(\phi_{2i-1})$ with each variable $x_j$
replaced by $x_{i,j}$.  Clearly,

\[\phi_{2i-1} \not \in \sat  \mbox{ iff } 
O_i \isoimpl 
\bigcup_{1 \leq j, \ell \leq n_i}  \{x_{i,j} \rightarrow x_{i,\ell}\},\]
where $n_i$ is the $n$ from $f(\phi_{2i-1})$.

For every $i$, $1 \leq i \leq k$, we define $E_i$ to be the set
of constraint applications $g(\phi_{2i})$ with each variable $y_j$
replaced by $y_{i,j}$.  Clearly,

\[\phi_{2i} \in \sat  \mbox{ iff } 
E_i \isoimpl \{y_{i,j} \vee y_{i,j+1} \ | \
1 \leq j < n_i'\},\]
where $n_i'$ is the $n$ from $g(\phi_{2i})$.

Note that the sets that occur to the right of $O_i \isoimpl$ are almost
isomorphic (apart from the number of variables).  The same
holds for the sets that occur to the right of $E_i \isoimpl$.  It is 
important to make sure that
these sets are exactly isomorphic. In order to 
do so, we simply pad the sets $O_i$ and $E_i$.

Let $n = \max\{n_i, n_i' + 2 \ | \ 1 \leq i \leq k\}$.
For $1 \leq i \leq k$, let
\[\widehat{O}_i = O_i \cup \{x_{i,1} \rightarrow x_{i,j},
x_{i,j} \rightarrow x_{i,1} \ | \ n_i < j \leq n\}.\]

$\widehat{O}_i$ is a set of constraint applications of ${\cal D}$,
since there exists a constraint application $A(x,y)$ of ${\cal D}$ that is
equivalent to $x \rightarrow y$
(see~\cite[Claim 14]{boe-hem-rei-vol:c:constraints}).
It is immediate that
\[\widehat{O}_i \isoimpl
\bigcup_{1 \leq j, \ell \leq n}  \{x_{i,j} \rightarrow x_{i,\ell}\}
\mbox{ iff }
{O}_i \isoimpl
\bigcup_{1 \leq j, \ell \leq n_i}  \{x_{i,j} \rightarrow x_{i,\ell}\}.\]

For $1 \leq i \leq k$, let
\[\widehat{E}_i = E_i \cup
\{y_{i,j} \vee y_{i,n_i'+1} \ | \  1 \leq j \leq n_i'\} \cup
\{y_{i,j} \vee y_{i,j+1} \ | \ n_i' + 1 \leq j < n\}.\]

Then
\[\widehat{E}_i \isoimpl \{y_{i,j} \vee y_{i,j+1} \ | \
1 \leq j < n\} \mbox{ iff }
E_i \isoimpl \{y_{i,j} \vee y_{i,j+1} \ | \
1 \leq j < n'_i\}.\]
The right-to-left direction is immediate.  The left-to-right
to direction can easily be seen if we think about this
as graphs.  Since $n \geq n_i' + 2$, any Hamiltonian path
in $\widehat{E}_i$ contains the subpath $n'_{i+1}, n'_{i+2}, \ldots, n$,
where $n$ is an endpoint.  This implies that there is a Hamiltonian
path in the graph restricted to $\{1, \ldots, n'_i\}$, i.e.,
in $E_i$.

So, our current situation is as follows.
For all $i$, $1 \leq i \leq k$,
$\widehat{O}_i$ is a set of constraint applications of ${\cal D}$ such that
\[\phi_{2i-1} \not \in \sat  \mbox{ iff } 
\widehat{O}_i \isoimpl 
\bigcup_{1 \leq j, \ell \leq n}  \{x_{i,j} \rightarrow x_{i,\ell}\}.\]
and 
$\widehat{E}_i$ is a set of constraint applications of $\lambda xy. x \vee y$
without duplicates such that
\[\phi_{2i} \in \sat  \mbox{ iff } 
\widehat{E}_i \isoimpl \{y_{i,j} \vee y_{i,j+1} \ | \
1 \leq j < n\}.\]

Our reduction is defined as follows

\[h(\phi_1, \ldots, \phi_{2k}) = \langle S,U \rangle,\]
where
\[S = \bigcup_{i=1}^k \left (\widehat{O}_{i} \cup \widehat{E}_{i} \cup
\bigcup_{1 \leq j,\ell \leq n} \{ x_{i,j}
\rightarrow y_{i,\ell}\} \right )\]
and
\[U = 
\bigcup_{1 \leq j, \ell \leq n}  \{x_{j} \rightarrow x_{\ell}\}   \cup
\bigcup_{j = 1}^{n-1} \{y_{j} \vee y_{j+1}\} \cup
\bigcup_{1 \leq j, \ell \leq n}  \{x_{j} \rightarrow y_{\ell}\}.  \]

Clearly, $h$ is computable in polynomial time and
$S$ and $U$ are sets of constraint applications of ${\cal C}$,
since there exists a constraint application $A(x,y)$ of ${\cal D}$ that is
equivalent to $x \rightarrow y$.

It remains to show that there exists an $i$ such that
$\widehat{O}_i \isoimpl \bigcup_{1 \leq j, \ell \leq n}
\{x_{i,j} \rightarrow x_{i,\ell}\}$ and
$\widehat{E}_i \isoimpl \{y_{i,j} \vee y_{i,j+1} \ | \
1 \leq j < n\}$ if and only if $S \isoimpl U$.

For the left-to-right direction, let $i_0$ be such that
$1 \leq i_0 \leq k$,  $\widehat{O}_{i_0} \isoimpl
\bigcup_{1 \leq j, \ell \leq n} \{x_{i_0,j} \rightarrow x_{i_0,\ell}\}$ and
$\widehat{E}_{i_0} \isoimpl \{y_{i_0,j} \vee y_{i_0,j+1} \ | \
1 \leq j < n\}$.
Let $\pi_x$ be a permutation of $\{x_{i_0,1}, \ldots, x_{i_0,n}\}$
such that $\pi_x(\widehat{O}_{i_0}) \Rightarrow
\bigcup_{1 \leq j, \ell \leq n} \{x_{i_0,j} \rightarrow x_{i_0,\ell}\}$ and
let $\pi_y$ be a permutation of $\{y_{i_0,1}, \ldots, y_{i_0,n}\}$
such that $\pi_y(\widehat{E}_{i_0})
\Rightarrow \{y_{i_0,j} \vee y_{i_0,j+1} \ | \ 1 \leq j < n\}$.

Define a permutation $\pi$ on the variables occurring in $S \cup U$ 
such that for all $1 \leq j,\ell \leq n$,
$\pi(x_{i_0,j}) = x_{\ell}$ if $\pi_x(x_{i_0,j}) = x_{i_0,\ell}$ and
$\pi(y_{i_0,j}) = y_{\ell}$ if $\pi_y(y_{i_0,j}) = y_{i_0,\ell}$.
It is immediate that 
$\pi(\widehat{O}_{i_0}) \Rightarrow \bigcup_{1 \leq j, \ell \leq n}
\{x_{j} \rightarrow x_{\ell}\}$,
$\pi(\widehat{E}_{i_0}) \Rightarrow \{y_{j} \vee y_{j+1} \ | \
1 \leq j < n\}$, and
$\pi(\bigcup_{1 \leq j, \ell \leq n}  \{x_{i_0,j} \rightarrow y_{i_0,\ell}\})
\Rightarrow 
\bigcup_{1 \leq j, \ell \leq n}  \{x_{j} \rightarrow y_{\ell}\}$.
It follows that $\pi(S) \Rightarrow U$.

For the converse, suppose that $S \isoimpl U$.
It is easy to see (see Lemma~\ref{l:variables})
that there exists a permutation $\pi$ of the variables 
that occur in $S \cup U$ such that $\pi(S) \Rightarrow U$ and such 
that for all $j$, $1 \leq j \leq n$, $\pi(x_j)$ and $\pi(y_j)$ do not 
occur in $U$.

We will now show that for all $1 \leq j \leq n$, $\pi$ cannot map a 
$y$-variable to $x_j$. For suppose that $\pi(y_{i,\ell}) = x_{j}$.
$S$ is satisfied by the assignment that sets all $y$-variables to 1
and all $x$-variables to 0, and $S$ remains satisfied if in this
assignment we change
the value of $y_{i,\ell}$ to 0 (recall that if $z \vee z' \in E_j$ then 
$z \neq z'$).
Then $\pi(S)$ is satisfied by the assignment that sets $\pi(y)$ to 1 for all
$y$-variables and $\pi(x)$ to 0 for all $x$-variables, and $\pi(S)$ is
still satisfied if in this assignment we change the value of 
$\pi(y_{i,\ell})$ to 0. But this is a contradiction, since $\pi(y_{i,\ell})
=x_j$ and 
changing the value of $x_j$ in a satisfying assignment for $U$ will always
make $U$ false.

Let $i_0,j_0$ be such that $\pi(x_{i_0,j_0}) = x_1$.
Now suppose that $\pi(z) = x_j$.  Then $z = x_{i,\ell}$.
Since $\pi(S) \Rightarrow (x_1 \leftrightarrow x_j)$,
$S \Rightarrow (x_{i_0,j_0} \leftrightarrow x_{i,\ell})$.
It follows that $i = i_0$, and thus,
$\pi(\{x_{i_0,\ell} \ | \ 1 \leq \ell \leq n \})$ =
$\{x_{\ell} \ | \ 1 \leq \ell \leq n \}$.

Next, suppose that $\pi(z) = y_j$. 
Since $\pi(S) \Rightarrow (x_{1} \rightarrow y_j)$,
$S \Rightarrow (x_{i_0,j_0} \rightarrow z)$.
It follows that $z = x_{i_0,\ell}$ or $z = y_{i_0,\ell}$.
Since $\pi(\{x_{i_0,\ell} \ | \ 1 \leq \ell \leq n \})$ =
$\{x_{\ell} \ | \ 1 \leq \ell \leq n \}$, the only possibility is
$z = y_{i_0,\ell}$.  It follows that
$\pi(\{y_{i_0,\ell} \ | \ 1 \leq \ell \leq n \})$ =
$\{y_{\ell} \ | \ 1 \leq \ell \leq n \}$.

Let $\alpha$ be the partial assignment that sets all $y$-variables to $1$,
and all $x$-variables except those in 
$\{x_{\ell} \ | \ 1 \leq \ell \leq n \}$ to 0.
Then $\pi(S)[\alpha]$ is equivalent to $\pi(\widehat{O}_{i_0})$ and
$U[\alpha]$ is equivalent to 
$\bigcup_{1 \leq j, \ell \leq n}  \{x_{j} \rightarrow x_{\ell}\}$.
Since $\pi(S) \Rightarrow U$, 
$\pi(S)[\alpha] \Rightarrow U[\alpha]$, i.e.,
$\pi(\widehat{O}_{i_0}) \Rightarrow 
\bigcup_{1 \leq j, \ell \leq n}  \{x_{j} \rightarrow x_{\ell}\}$,
and thus $\widehat{O}_{i_0} \isoimpl 
\bigcup_{1 \leq j, \ell \leq n}  \{x_{i_0,j} \rightarrow x_{i_0,\ell}\}$.

Let $\beta$ be the partial assignment that sets all $x$-variables to $0$,
and all $y$-variables except those in
$\{y_{\ell} \ | \ 1 \leq \ell \leq n \}$ to 1.
Then $\pi(S)[\beta]$ is equivalent to $\pi(\widehat{E}_{i_0})$ and
$U[\beta]$ is equivalent to 
$\bigcup_{j = 1}^{n-1} \{y_{j} \vee y_{j+1}\}$.
Since $\pi(S) \Rightarrow U$, $\pi(S)[\beta] \Rightarrow U[\beta]$.
It follows that $\pi(\widehat{E}_{i_0}) \Rightarrow 
\bigcup_{j = 1}^{n-1} \{y_{j} \vee y_{j+1}\}$, and thus
$\widehat{E}_{i_0} \isoimpl 
\bigcup_{j = 1}^{n-1} \{y_{i_0,j} \vee y_{i_0,j+1}\}$.
This completes the proof of Theorem~\ref{t:parallelnphard}.
\end{proof}

It should be noted that constructions similar to the proof
of Theorem~\ref{t:parallelnphard} can be used to prove
$\parallelnp$-hardness for some other cases as well.
However, new insights and constructions will be needed
to obtain $\parallelnp$-hardness for all non-Schaefer cases.

\section{Open Problems}

The most important question left open by this paper is
whether Conjecture~\ref{c:isoimp} holds.  In addition,
the complexity of the isomorphic implication problem
for Boolean formulas is still open. This problem is
trivially in $\Sigma_2^p$, and, by Theorem~\ref{t:parallelnphard},
$\parallelnp$-hard.  Note that an improvement of the upper
bound will likely give an improvement of the best-known
 upper bound ($\Sigma_2^p$) for the isomorphism problem for 
Boolean formulas, since that problem is 2-conjunctive-truth-table
reducible to the isomorphic implication problem.

Schaefer's framework is not the only framework to study
generalized Boolean problems.  It would be interesting
to study the complexity of isomorphic implication in other
frameworks, for example, for Boolean circuits over a
fixed base.

\medskip
\noindent \textbf{Acknowledgments:}
The authors thank Henning Schnoor and Heribert Vollmer for 
helpful comments.

\goodbreak
 {\small \bibliography{edith}}

\begin{thebibliography}{KSTW01}

\bibitem[AT00]{agr-thi:j:boolean-isomorphism}
M.~Agrawal and T.~Thierauf.
\newblock The formula isomorphism problem.
\newblock {\em SIAM Journal on Computing}, 30(3):990--1009, 2000.

\bibitem[BCRV04]{boe-cre-rei-vol:j:blocks-two}
E.~B{\"{o}}hler, N.~Creignou, S.~Reith, and H.~Vollmer.
\newblock Playing with {B}oolean blocks, part {II}: {C}onstraint {S}atisfaction
  {P}roblems.
\newblock {\em SIGACT News}, 35(1):22--35, 2004.

\bibitem[BH91]{bus-hay:j:tt}
S.~Buss and L.~Hay.
\newblock On truth-table reducibility to {SAT}.
\newblock {\em Information and Computation}, 91(1):86--102, 1991.

\bibitem[BHRV02]{boe-hem-rei-vol:c:constraints}
E.~B{\"{o}}hler, E.~Hemaspaandra, S.~Reith, and H.~Vollmer.
\newblock Equivalence and isomorphism for {B}oolean constraint satisfaction.
\newblock In {\em Proceedings of the 16th Annual Conference of the EACSL (CSL
  2002)}, pages 412--426. Springer-Verlag {\it Lecture Notes in Computer
  Science \#2471}, September 2002.

\bibitem[BHRV03]{boe-hem-rei-vol:t:con-iso-revised}
E.~B{\"{o}}hler, E.~Hemaspaandra, S.~Reith, and H.~Vollmer.
\newblock The complexity of {B}oolean constraint isomorphism.
\newblock Technical Report cs.CC/0306134, Computing Research Repository,
  \mbox{http://www.acm.org/repository/}, June 2003.
\newblock Revised, April 2004.

\bibitem[BHRV04]{boe-hem-rei-vol:c:iso}
E.~B{\"{o}}hler, E.~Hemaspaandra, S.~Reith, and H.~Vollmer.
\newblock The complexity of {B}oolean constraint isomorphism.
\newblock In {\em Proceedings of the 21st Symposium on Theoretical Aspects of
  Computer Science}, pages 164--175. Springer-Verlag {\it Lecture Notes in
  Computer Science \#2996}, March 2004.

\bibitem[BKJ00]{bul-jea-kro:c:finite-algebras}
A.~Bulatov, A.~Krokhin, and P.~Jeavons.
\newblock Constraint satisfaction problems and finite algebras.
\newblock In {\em Proceedings of the 27th International Colloquium on Automata,
  Languages and Programming}, pages 272--282. Springer-Verlag, 2000.

\bibitem[BR93]{bor-ran:t:subfunction}
B.~Borchert and D.~Ranjan.
\newblock The ciruit subfunction relations are {$\Sigma_2^p$}-complete.
\newblock Technical Report MPI-I-93-121, MPI, Saarbr{\"u}cken, 1993.

\bibitem[BRS98]{bor-ran-ste:j:boolean-equivalence}
B.~Borchert, D.~Ranjan, and F.~Stephan.
\newblock On the computational complexity of some classical equivalence
  relations on {B}oolean functions.
\newblock {\em Theory of Computing Systems}, 31(6):679--693, 1998.

\bibitem[CH96]{cre-heb:j:counting}
N.~Creignou and M.~Hermann.
\newblock Complexity of generalized satisfiability counting problems.
\newblock {\em Information and Computation}, 125:1--12, 1996.

\bibitem[CKS01]{cre-kha-sud:b:con}
N.~Creignou, S.~Khanna, and M.~Sudan.
\newblock {\em Complexity {C}lassifications of {B}oolean {C}onstraint
  {S}atisfaction {P}roblems}.
\newblock Monographs on Discrete Applied Mathematics. SIAM, 2001.

\bibitem[Coo71]{coo:c:theorem-proving}
S.~Cook.
\newblock The complexity of theorem-proving procedures.
\newblock In {\em Proceedings of the 3rd ACM Symposium on Theory of Computing},
  pages 151--158. ACM Press, 1971.

\bibitem[Cre95]{cre:j:max-sat}
N.~Creignou.
\newblock A dichotomy theorem for maximum generalized satisfiability problems.
\newblock {\em Journal of Computer and System Sciences}, 51:511--522, 1995.

\bibitem[GJ79]{gar-joh:b:int}
M.~Garey and D.~Johnson.
\newblock {\em Computers and Intractability: A Guide to the Theory of
  NP-Completeness}.
\newblock {W. H. Freeman and Company}, 1979.

\bibitem[Hem04]{hem:t:con-ph}
E.~Hemaspaandra.
\newblock Dichotomy theorems for alternation-bounded quantified {B}oolean
  formulas.
\newblock Technical Report cs.CC/0406006, Computing Research Repository,
  \mbox{http://www.acm.org/repository/}, June 2004.

\bibitem[HHR97]{hem-hem-rot:j:raising-lower-bounds}
E.~Hemaspaandra, L.~Hemaspaandra, and J.~Rothe.
\newblock Raising {NP} lower bounds to parallel {NP} lower bounds.
\newblock {\em SIGACT News}, 28(2):2--13, 1997.

\bibitem[JCG97]{jea-coh-gys:j:closure-properties-constraints}
P.~Jeavons, D.~Cohen, and M.~Gyssens.
\newblock Closure properties of constraints.
\newblock {\em Journal of the ACM}, 44(4):527--548, 1997.

\bibitem[Jea98]{jea:j:algebraic-structure}
P.~Jeavons.
\newblock On the algebraic structure of combinatorial problems.
\newblock {\em Theoretical Computer Science}, 200(1-2):185--204, 1998.

\bibitem[Jub99]{jub:c:usat}
L.~Juban.
\newblock Dichotomy theorem for generalized unique satisfiability problem.
\newblock In {\em Proceedings of the 12th Conference on Fundamentals of
  Computation Theory}, pages 327--337. Springer-Verlag {\it Lecture Notes in
  Computer Science \#1684}, 1999.

\bibitem[KK01]{kir-kol:c:minimal-sat}
L.~Kirousis and P.~Kolaitis.
\newblock The complexity of minimal satisfiability problems.
\newblock In {\em Proceedings of the 18th Symposium on Theoretical Aspects of
  Computer Science}, pages 407--418. Springer-Verlag {\it Lecture Notes in
  Computer Science \#2010}, 2001.

\bibitem[KS98]{kav-sid:j:inv-sat}
D.~Kavvadias and M.~Sideri.
\newblock The inverse satisfiability problem.
\newblock {\em SIAM Journal on Computing}, 28(1):152--163, 1998.

\bibitem[KST93]{koe-sch-tor:b:graph-iso}
J.~K{\"{o}}bler, U.~Sch{{\"{o}}}ning, and J.~Tor{\'{a}}n.
\newblock {\em The Graph Isomorphism Problem: Its Structural Complexity}.
\newblock Birkh{\"{a}}user, 1993.

\bibitem[KSTW01]{kha-sud-tre-wil:j:approx-constraints}
S.~Khanna, M.~Sudan, L.~Trevisan, and D.~Williamson.
\newblock The approximability of constraint satisfaction problems.
\newblock {\em SIAM Journal on Computing}, 30(6):1863--1920, 2001.

\bibitem[Pos44]{pos:j:re}
E.~Post.
\newblock Recursively enumerable sets of integers and their decision problems.
\newblock {\em Bulletin of the AMS}, 50:284--316, 1944.

\bibitem[Sch78]{sch:c:sat}
T.~Schaefer.
\newblock The complexity of satisfiability problems.
\newblock In {\em Proceedings of the 10th ACM Symposium on Theory of
  Computing}, pages 216--226, 1978.

\bibitem[Wag87]{wag:j:more-on-bh}
K.~Wagner.
\newblock More complicated questions about maxima and minima, and some closures
  of {NP}.
\newblock {\em Theoretical Computer Science}, 51(1--2):53--80, 1987.

\end{thebibliography}

\end{document}